\documentclass[letterpaper, 10 pt, conference]{ieeeconf}  

\IEEEoverridecommandlockouts                              

\usepackage{amsmath} 
\usepackage{amssymb}  
\usepackage{subcaption}
 
\usepackage[short]{optidef}
\usepackage{pgf}
\usepackage[utf8x]{inputenc}
\usepackage{cite}
\usepackage{hyperref}

\title{\LARGE \bf
Reference design for closed loop system optimization
}

\author{Samuel Balula$^{1,2}$, Alex Liniger$^{3}$, Alisa Rupenyan$^{1,2}$ and John Lygeros$^{1}$
\thanks{$^{1}$The authors are with the Automatic Control Laboratory, Department of Electrical Engineering and Information Technology, ETH, Zurich, Switzerland.
        {\tt\small \{balula, ralisa, jlygeros\}@control.ee.ethz.ch}}
\thanks{$^{2}$The authors are with Inspire AG, ETH, Zurich, Switzerland.
        }%
\thanks{$^{3}$The author is with the Computer Vision Lab, Department of Electrical Engineering and Information Technology, ETH, Zurich, Switzerland.
        {\tt\small alex.liniger@vision.ee.ethz.ch}}
}

\begin{document}

\maketitle

\begin{abstract}
An optimization-based method for improving the productivity of precision machine tools is proposed, where the reference path is computed in local coordinates, and information about the machine tool performance is learned from experimental data. The optimization yields a modified reference that is tracked by the existing low-level controller. The method is tested in simulation for a biaxial positioning system. The positioning system is modelled as double integrator, and the controller characteristic is modelled from experimental data using a least-squares fit. Simulation results show that the method is effective in designing optimal references even for challenging geometries such as sharp corners. The application of this procedure allows the retrofit of the control of existing machines with minimal overhead, by providing a modified reference file to track.
\end{abstract}

\section{Introduction}
Control of high-precision positioning systems aims to achieve tracking of a geometrical path with micrometer accuracy while maintaining speeds above few meters per second. Precise modeling of the machining system enables the application of predictive control methods to achieve high tracking accuracy. Of particular interest is the work done by Lam et al \cite{MPCC}, where the contour, defined as the desired geometry to be traversed, is parametrized using the arc-length of the reference path. A non-linear Model Predictive Control (MPC) formulation is proposed which trades off the contouring error and the traversal speed. One of the challenges in MPC-based methods is the time needed to perform optimization, which should be within the sampling time of the system, often in the order of 1 ms. We have recently proposed a method that achieves sub-ms computational time while maintaining a very high tracking accuracy, using local coordinate system to formulate the tracking problem \cite{Liniger2019}, demonstrating it on a simple model of the system. Similar to driving along a racing track, a linear time-varying problem is linearized and approximated with a quadratic program, to enable fast solving times \cite{kabzan2019learning}. Here we build on that method, with the goal of accounting for non-linearities and suboptimal low-level machine performance through learning patterns in the machine data and using them to provide an optimized reference path.

More and more often, due to the scaling up of manufacturing operations, systems have to be designed or retrofitted to achieve a diverse set of objectives, to be operated in parallel with other systems, and for high-precision machining, to achieve more complex geometries. To comply with these requirements considering the restricted access to the control algorithm, methods that allow modification of the reference or the set points given to the controller system can be of practical use, learning the performance of the machine from samples and using optimization-based predictive control algorithms to provide new, optimized reference without changing the underlying low-level cascaded PID control. An established approach that modifies the reference command to a closed-loop system to satisfy constraints in the systems is known as reference-governor controller \cite{RefGTutorial}. The primary purpose of the method is to provide constraints satisfaction, as stability and tracking performance is accomplished by the low-level controllers already in place. It provides optimal set points (reference) for the primary control layer, using optimization-based high-level controller (governor), where dynamic constraints are included. In discrete vector reference-governor, the modified reference is calculated using quadratic programming techniques, for each time instance. To avoid the iterative computations, the authors suggest to use explicit multi-parametric quadratic programming \cite{Garone}. Extending the method further by including soft constraints and a specific structure in the formulation of the space-state model of the system brings the method fully to a model predictive control (MPC) formulation \cite{MPCRefGov}. In many cases, knowledge of the low-level controller structure is required to implement reference or command governors.

Regarding the modification of the reference, a successful strategy used in manufacturing and process control is to take advantage of the repetitive structure of the process and to correct for the performance error in a learning phase. Iterative learning control (ILC) provides a framework that has been extensively explored for manufacturing problems \cite{ILCsurvay, ILChighPres, mandra2018performance}. Repetitive production operations provide the opportunity to use sensing and actuation monitoring to learn the effects of exogenous disturbances and complex dynamics on the process. Machine wear and feedstock variation are two typical manufacturing examples that require process control adaptation in order to maintain quality and throughput. Iterative Learning Control adapts the feed forward commands to the process machinery to compensate for the effects of such disturbances \cite{ILCquadrotors, ILC_plasma}. In many cases, ILC requires access to the controller of the system for an efficient implementation.

The challenge of developing an accurate system model to apply MPC techniques is overcome in learning control, where black-box models replace the mathematical models of the plant and the controllers, or first principles models are complemented by learning the deviations from the model \cite{LearnControl1, GPcontrol, GProbotics1, GPlearningControl}. Often, probabilistic modeling is preferred in the learning of the system dynamics, as it provides the associated uncertainty with the predictions, which can be included in the constraints of the underlying MPC-based controller \cite{GPcars}. A simple nominal model is often available in practice, while more complex nonlinearities can be challenging and time-intensive to model from first principles. Model predictive control approach that integrates this nominal system with an additive part of the dynamics modeled as a Gaussian process (GP), where the resulting nonlinear stochastic control problem takes into account the model uncertainties associated with the GP has been applied in the field of robotics (see for example \cite{GProb2, RobLearn4, roboticsGP5}), but to our knowledge, there are no demonstrations or adaptations to manufacturing and machining applications. Often resulting algorithms are showcased on available systems with flexible architecture, such as quadrotors \cite{QuadrGP}, bipedal or quadrupedal robots \cite{humanoidGP}. While these methods are promising, their real-time implementation requires either access to the controller of the system, or large interventions in the system architecture. Non-parametric learning models as Gaussian process regression require the availability of strong computational capabilities.

We present a method that address adaptability and performance of high-precision machining systems, taking advantage of their high repeatability and mechanical stability. The adaptation is done in a learning phase offline where the system and its built-in controller are considered as one block, and the reference provided to that block is modified to achieve high performance. The repeatability of such systems enables the relaxation of causality in the control method, and learning of the machine performance without an advanced control method combined with optimization can be applied offline to enhance high tracking performance and adaptability. 

This paper is organized as follows: In section \ref{sec:method} the method is described, in particular in \ref{ssec:problem-formulation} with the definition of the problem, relevant variables in \ref{ssec:variables}, the coordinates transformation to local reference frames in \ref{ssec:local}, the double integrator model in \ref{ssec:model}, and the statement of the optimization problem in \ref{ssec:opt}.  
The results are compiled in \ref{sec:results}, followed by the conclusion \ref{sec:conclusion}.

\section{Method}
\label{sec:method}
\subsection{Problem formulation}
\label{ssec:problem-formulation}

In this paper we aim to improve the performance of a system, without introducing modifications to the low-level control loop structure, or in any of the involved control parameters. The problem of interest is a two-dimensional machine tool contouring control problem, with the goal of finishing a predefined path as fast as possible, given a tolerance band in the order of few tens of micrometers.
For the two-dimensional positioning system a lumped mass model is used where the acceleration in each dimension can be controlled individually. The resulting model is a double integrator, with bounds on the states and inputs due to mechanical constraints and contour tolerances. The mismatch between the output of this simple model, and the experimental machine output in tracking the predefined path is learned from experimental data.
This approach enables an increase in performance without requiring changes in the control architecture or in the machine design.

\subsection{Variables}
\label{ssec:variables}
The 2D system is split in two independent axis, $x$ and $y$. We introduce the state
\begin{equation}
	\omega(k) = \begin{bmatrix}
		x_\omega(k)		    &y_\omega(k)	&1\\
		\dot x_\omega(k)	&\dot y_\omega(k) &0
	\end{bmatrix}\mathrm{,}
\end{equation}
and input
\begin{equation}
    u(k) = \begin{bmatrix}
        u_x(k)  &u_y(k)  &0\\
        \end{bmatrix}\mathrm{,}
\end{equation}
where, to ease coordinate transformations, the state is written as a matrix instead of a vector along with the 3\textsuperscript{rd} column $\{0, 1\}$ entries. This structure imposes independence of the axis, as discussed in \ref{ssec:opt}.

We introduce two additional variables, with similar structure as the state $\omega$ and input $u$ defined above: the reference $\gamma$ and virtual input $v$.
\if 01
\begin{equation}
	\gamma(k) = \begin{bmatrix}
		x_\gamma(k)		    &y_\gamma(k)	&1\\
		\dot x_\gamma(k)	&\dot y_\gamma(k) &0
	\end{bmatrix}\mathrm{,}
\end{equation}
\begin{equation}
    v(k) = \begin{bmatrix}
        v_x(k)  &v_y(k)  &0\\
        \end{bmatrix}\mathrm{.}
\end{equation}
\fi
We denote the objective (target contour) as
\begin{equation}
    \xi(k) = \begin{bmatrix}
        x_\xi(k)    &y_\xi(k)   &\alpha(k)\\
    \end{bmatrix}\mathrm{,}
\end{equation}
where $\alpha(k)$ is the orientation of the tangent to each point in the shape,
and finally the time and non-constant time step 
\begin{equation}
    t(k+1) = t(k) + \Delta t(k)\mathrm{.}
\end{equation}

The variables introduced can be defined as follows:

The objective $\Xi = \{\xi(k): k\in\{1,\dots,N\}\}$, where $N$ is the number of points that define the trajectory, is the shape or contour that we are interested in tracking.
It is defined in 2D space without any information about time, velocities or accelerations. The deviations are to be evaluated perpendicular to the orientation $\alpha(k)$.
Although $\Xi$ may be defined with an arbitrarily large number of points, in practice a computationally tractable number is used.
A simple choice is to sample the idealized contour at constant space steps.

The reference $\Gamma = \{(t(k), \gamma(k)): k\in\{1,\dots,N\}\}$ is the set of positions and velocities that the machine is given, along with time information for each point.

The output $\Omega = \{(t(k), \omega(k)): k\in\{1,\dots,N\}\}$ is the expected trajectory produced by the machine when given the reference $\Gamma$. The output time steps are the ones defined by the reference. In this work we assume full observability of the system.

\if 01

\begin{figure}[t]
  \begin{center}
    \includegraphics[width=.9\columnwidth]{img/m1xy1} 
    \caption{Scheme of the optimization variables, where $(x_\xi, y_\xi)$ denotes the objective (target contour), $(x_\gamma, y_\gamma)$ the position components of the reference, $(x_\omega, y_\omega)$ the output, $(e_x, e_y)$ the error, $(u_x, u_y)$ the input (scaled) and $(\dot x_\omega, \dot y_\omega)$ is the output velocity (scaled) on a 2D coordinate system.
    }
    \label{fig:m1}
  \end{center}
\vspace{-0.5cm}
\end{figure}
\fi

\begin{figure}[t]
  \begin{center}
    \includegraphics[width=.9\columnwidth]{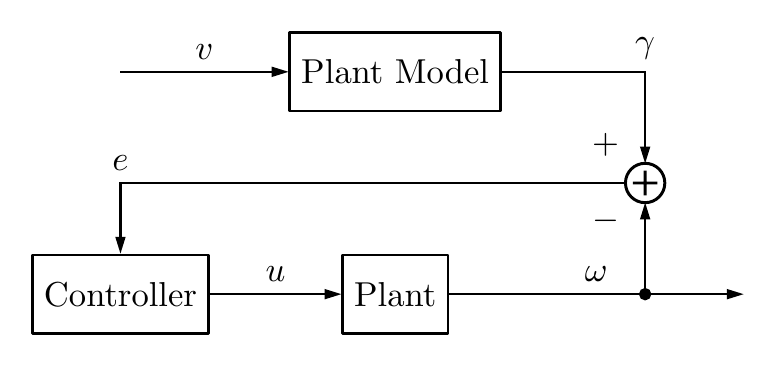} 
    \caption{Control loop schematic, where $\gamma$ denotes the reference, $\omega$ the output, $e$ the error, $u$ the input and $v$ the virtual input. During the optimization phase, the controller and plant are replaced by their models. On the experimental phase, the precomputed reference $\gamma$ is read from a file.
    }
    \label{fig:variables}
  \end{center}
	\vspace{-0.5cm}
\end{figure}

\subsection{Coordinate transformation}
\label{ssec:local}
For each point $k$ of the objective a {\it local} reference frame is defined. It has the origin at $(x_\xi(k), y_\xi(k))$ and orientation $\alpha(k)$.

The state $\omega(k)$ and virtual state $\gamma(k)$ can be written both in the global frame or in the local coordinate frame. The transformation between the two coordinates can be easily computed with
\begin{equation}
	\omega^g(k) = \omega(k) T^g_k\mathrm{,}
\end{equation}
where
\begin{equation}
	T^g_k = \begin{bmatrix}
		\cos \alpha(k)		&\sin \alpha(k)	&0\\
		-\sin \alpha(k)		&\cos \alpha(k)	&0\\
		x_\xi(k) 	        &y_\xi(k)	    &1\\
	\end{bmatrix}\mathrm{,}
\end{equation}

Assume that the control input $u$ of the system, for which a simple model is available, can be described as a function of the current error.
Fig. \ref{fig:variables} shows the control loop scheme.
This corresponds to a PD controller, i.e. the input is modeled as a linear function of the position and velocity error. This equation can be modified, extending the method for other controller types
Fig. \ref{fig:variables} shows the control loop scheme.

\begin{equation}
\label{eq:u-fit-lin}
u_j(k+1) = \lambda_j' e_j(k) = \lambda_j' (\gamma_j^g(k) - \omega_j^g(k)), ~j=\{x, y\}\mathrm{,}
\end{equation}
where $e$ is the error, the reference $\gamma^g(k)$ and output $\omega^g(k)$ are both written global coordinates $g$, and
\begin{equation}
\lambda = \begin{bmatrix}
K_x^P   &K_y^P\\[6pt]
K_x^D   &K_y^D\\
\end{bmatrix}\mathrm{,}
\end{equation}
where each column of $\lambda \in \mathbb{R}^{2 \times 2}$ contains the proportional $K_j^P$ and derivative $K_j^D$ gains.

The model mismatch and the controller characteristics can be identified from experimental data and provided through \eqref{eq:u-fit-lin}.
Data collected from the available $(x,y)$ positioning system with industrial grade actuators and sensors~\cite{Lanz_2018ijat} can be used to learn the relation \eqref{eq:u-fit-lin} between the accelerations, used as inputs, and the tracking performance errors.


\subsection{Model}
\label{ssec:model}
The state space differential equation that describes the time evolution of the system state, assuming a linear time invariant (LTI) model in continuous time is 

\begin{equation}
	\dot \omega^g(t) = A^c \omega^g(t) + B^c u_\omega(t)\mathrm{.}
\end{equation}
For variables $\omega$ and $\gamma$ we denote global coordinates with$~^g$, otherwise they are written in local coordinates. Using standard integration the equivalent discrete time system can be written as
\begin{equation}
\begin{split}
    \label{eq:int}
	\omega^g(t(k+1)) & = e^{A^c \Delta t(k)} \omega^g(t(k)) \\
	&+ \int_0^{\Delta t(k)} e^{A^c (\Delta t(k) - \tau)} B^c u(\tau) d\tau	\mathrm{,}
	\end{split}
\end{equation}
or equivalently, with zero order hold of the input
\begin{equation}
	\label{discrete-sys}
	\omega^g(k+1) = A(k) \omega^g(k) + B(k) u(k)\mathrm{.}
\end{equation}

The system dynamics can be written in local coordinates
\begin{equation}
	\omega(k+1) = A(k)~\omega(k) T^g_k T^{k+1}_g + B(k)~u(k) T^{k+1}_g {,}
\end{equation} 
where $T^{k+1}_g = {(T^g_{k+1}})^{-1}$.
\vspace{.2cm}

For simplicity a double integrator model for the open loop system and a PD controller are assumed for each axis. Other structures and dynamics can be handled in a similar way.
\begin{equation}
	A(k) = \begin{bmatrix}
		1 &\Delta t(k)\\
		0 &1
	\end{bmatrix}\mathrm{,}
\end{equation}

\begin{equation}
\label{dt2}
	B(k) = \begin{bmatrix}
		\frac{1}{2}{\Delta t(k)}^2\\
		\Delta t(k)
	\end{bmatrix}\mathrm{.}
\end{equation}
We note, that this transformation corresponds to the evaluation of deviations and velocities at specific points of the objective, by reading directly the local coordinates. The time interval $\Delta t(k)$ {between output points $k$ and $k+1$} is not constant, but corresponds to constant intervals in the space discretization of the objective.

It should be pointed out that the low level controller operates with a constant time step. Usually it is much smaller ($\approx 100~\mathrm{\mu s}$) than the one used in the optimization, where the number of points is kept small for a fast ($\approx 60~\mathrm{s}$) computation of the reference. Note that this is an offline computation for the whole contour, unlike the receding horizon based approach in \cite{Liniger2019}. Additionally, since the reference has varying time steps, a transformation must be performed.
One approach to translate the optimal solution to a more realistic scenario is to take a constant $\Delta t$ in \eqref{eq:int}, use the optimized variables $\{(t(k), v(k)): k\in\{1,\dots,N\}\}$ and compute $\Gamma$ with the desired sample rate.
This procedure may result in a small loss of precision. The study of such effects is the subject of future work.

\subsection{Optimization problem}
\label{ssec:opt}
Following the goal described in \ref{ssec:problem-formulation}, the cost function can be written as
\begin{equation}
	\begin{split}
		J & = \sum_{k=1}^{N} \Delta t(k)\\
		& +\sum_{j={x,y}} \sum_{k=1}^{N}{\omega_j^{T}(k) Q_\omega \omega_j(k)}
		+\sum_{k=1}^{N-1} u(k) R_u u^T(k) \\
		& +\sum_{j={x,y}} \sum_{k=1}^{N-1}{\gamma_j^{T}(k) Q_\gamma \gamma_j(k)}
		+\sum_{k=1}^{N-2} v(k) R_v v^T(k)\mathrm{,}
		\end{split}
\end{equation}
where $Q_\omega$, $R_u$, $Q_\gamma$, and $R_v$ are positive definite weight matrices. The term on $\Delta t$ penalizes the amount of time it takes to follow the trajectory, while the term in $\omega$ penalizes the deviation from the objective, and the regularization term in the input $u$, promotes smooth trajectories.

Additionally we add two artificial terms that depend on the virtual input $v$, whose role it is to smooth the reference $\Gamma$, and a term on $\gamma$ that penalizes deviations of the reference. By smoothing and penalizing the distance between reference and objective, exploitation of references with high frequency oscillations is prevented.

Note that the cost function is not minimal, in the sense that there are several terms that contribute for the same effect as for example time and tangential speed. We choose to keep all these terms for ease of tuning, while for online implementation, the cost function needs to be simplified further.

The system model is enforced in the constraints, as well as the input dependency in the error


\if 01
\begin{align}
\min_{\mathbf{\omega},\mathbf{u}} \; \quad J& \nonumber\\
\text{s.t} \quad\; &{x_\omega(k)}{ = 0}\nonumber\\
&{\omega(k+1)}{ = A \omega(k) T^g_k T^{k+1}_g + B u(k) T^{k+1}_g} \nonumber\\
&{\gamma(k+1)}{ = A \gamma(k) T^g_k T^{k+1}_g + B v(k) T^{k+1}_g} \nonumber\\
&{\omega(k)}{\in \Omega}, \quad v \in \mathcal{V}, \quad u \in \mathcal{U}, \quad {\Delta t(k)}{\in \mathbb{R}^+} \nonumber\\
&{u(k)}{ = \lambda (\gamma^g(k) - \omega^g(k))}\nonumber\\
&k=0,..,N {,}
\label{eq:fullOpt}
\end{align}
\fi

\begin{mini}{}{J}{}{}
	\addConstraint{\omega(k+1)}{ = A(k)~ \omega(k) T^g_k T^{k+1}_g + B(k)~ u(k) T^{k+1}_g}{}
	\addConstraint{\gamma(k+1)}{ = A(k)~ \gamma(k) T^g_k T^{k+1}_g + B(k)~ v(k) T^{k+1}_g}{}
	\addConstraint{u_j(k+1)}{ = \lambda_j' (\gamma_j^g(k) - \omega_j^g(k))}{~ \quad \quad \quad j=\{x, y\}}
	\addConstraint{x_\omega(k)}{ = 0}{}
	\addConstraint{u(k)}{ \in \mathcal{U}}{}
	\addConstraint{\omega(h)}{\in \mathcal{W}}{}
	\addConstraint{\Delta t(k)}{\in \mathbb{R}^+}{}
	\addConstraint{k }{= \{1, \dots, N\}}{}
	\addConstraint{h }{= \{16, \dots, N\}}{}
    \label{eq:fullOpt}
    \mathrm{,}
\end{mini}
where the constraint on $x_\omega(k)$ ensures that the evaluation of the deviation is done perpendicular to the $k$\textsuperscript{th} point of the objective $\xi(k)$. This constraint is also key to the time interval $\Delta t(k)$, as it effectively fixes the tangential space distance between two consecutive points of the objective.
\begin{equation}
\label{bounds}
\begin{split}
\mathcal{U}  = \{(u_x, u_y) \in \mathbb{R}: &|u_x| \leq 2~m~s^{-2}\wedge  |u_y| \leq 2~m~s^{-2}\}\mathrm{,}\\
\mathcal{W}  = \{\omega \in \mathbb{R}:  &|\dot x_\omega| \leq 2~m~s^{-1} \wedge  |\dot y_\omega| \leq 2~m~s^{-1}\\
 &\wedge |y_\omega| \leq 20~\mu m\}\mathrm{.}
\end{split}
\end{equation}
Notice that no constraints are imposed on $v$ and $\gamma$.
The structure of the optimization variables and cost function decouples, by design, the states of each axis. Therefore no cross terms are present on the cost function.
The resulting optimization objective is quadratic with bilinear constraints. The bilinearity arises due to the product of $\Delta t$ with velocities and ${\Delta t}^2$ with the input, see \eqref{dt2}.
The constraints are not enforced in the first points of the output. This is consistent with the usual practice of machine tools to have a small initial path before starting with the production of the actual part. 

\section{Results}
\label{sec:results}

\subsection{Closed loop identification}

\begin{figure}[t]
  \begin{center}
	  \vspace{-0.3cm}
    \includegraphics[width=.9\columnwidth]{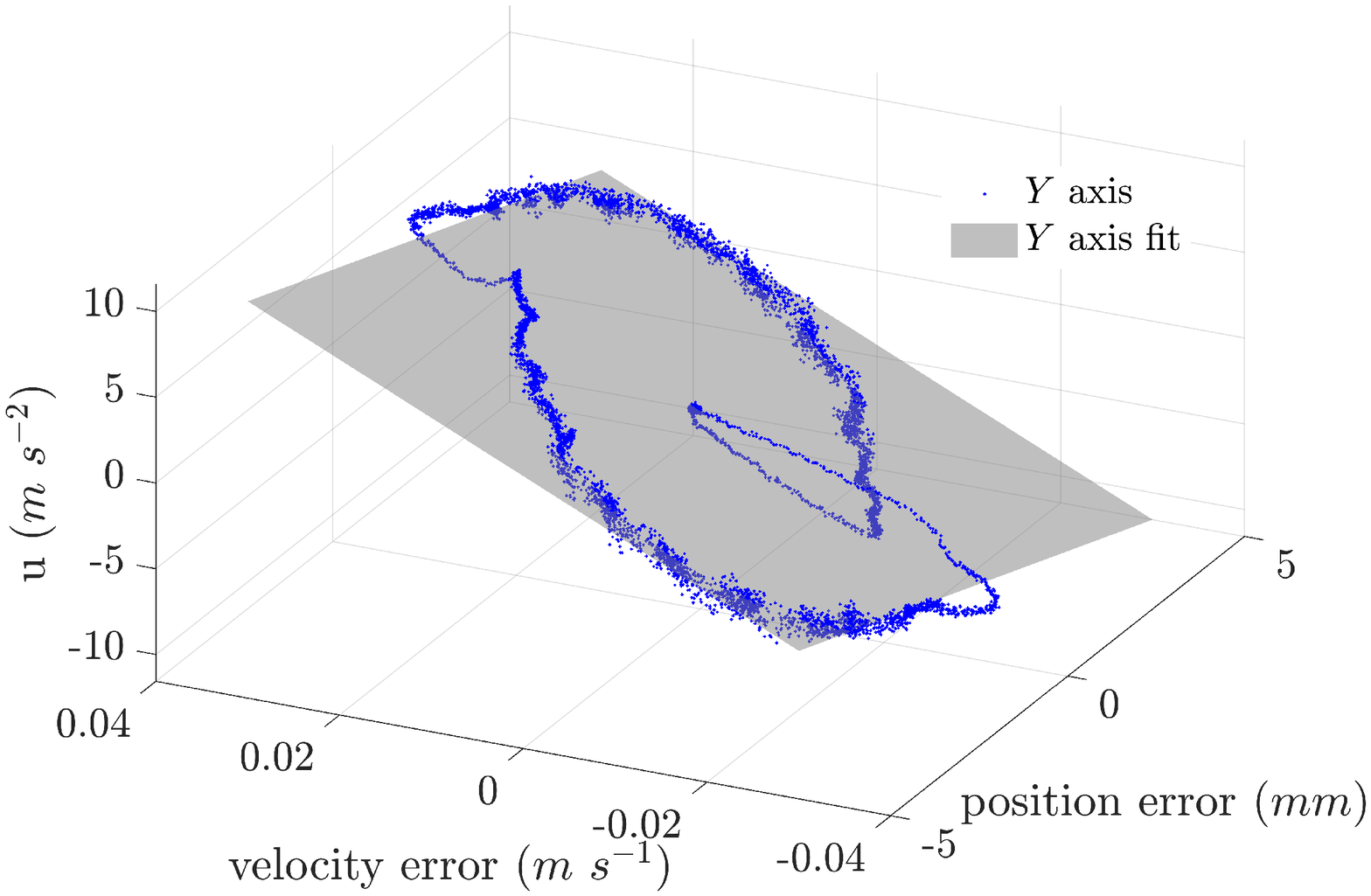}
	  \caption {Typical experimental input $u$ as a function of position and velocity errors shown for the $y$-axis, when tracking a circular path. Linear least square fit of the experimental data represented with a plane. A similar plot can be produced for the $x$ axis. The computation time for the fit is $<1\mathrm{s}$.}
   \label{fig:3dfit} 
  \end{center}
\vspace{-0.5cm}
\end{figure}

Data collected from the experimental apparatus has been used with \eqref{eq:u-fit-lin} to identify $\lambda$. The least squares fit between the experimentally obtained errors in position and velocity, and the accelerations is shown in Fig. \ref{fig:3dfit} for the $y$ axis. The $x$ axis exhibits a similar behavior. We note that the repeatability of the system is better than the encoder accuracy of $3~\mathrm{\mu m}$.
The identified $\lambda$ is then used in the rest of this work. The results shown are simulations of a discrete time double integrator with varying sampling time.

\subsection{Reference generation}

The nonlinear optimization problem \eqref{eq:fullOpt} has been solved using {\it IPOPT} v3.12 and the programming language {\it Julia} v1.2. 
The weights used in the optimization are $Q_\omega = \mathrm{diag}(10^7, 1)$, $R_u=I_2$, $Q_\gamma = \mathrm{diag}(10^6, 1)$, $R_v = I_2$. The high values for the deviation are a consequence of the order of magnitude of this quantity that typically lies in the order of $10^{-5}$.

\begin{figure}[t]
	\vspace{3mm}
  \begin{center}
    \includegraphics[width=.92\columnwidth]{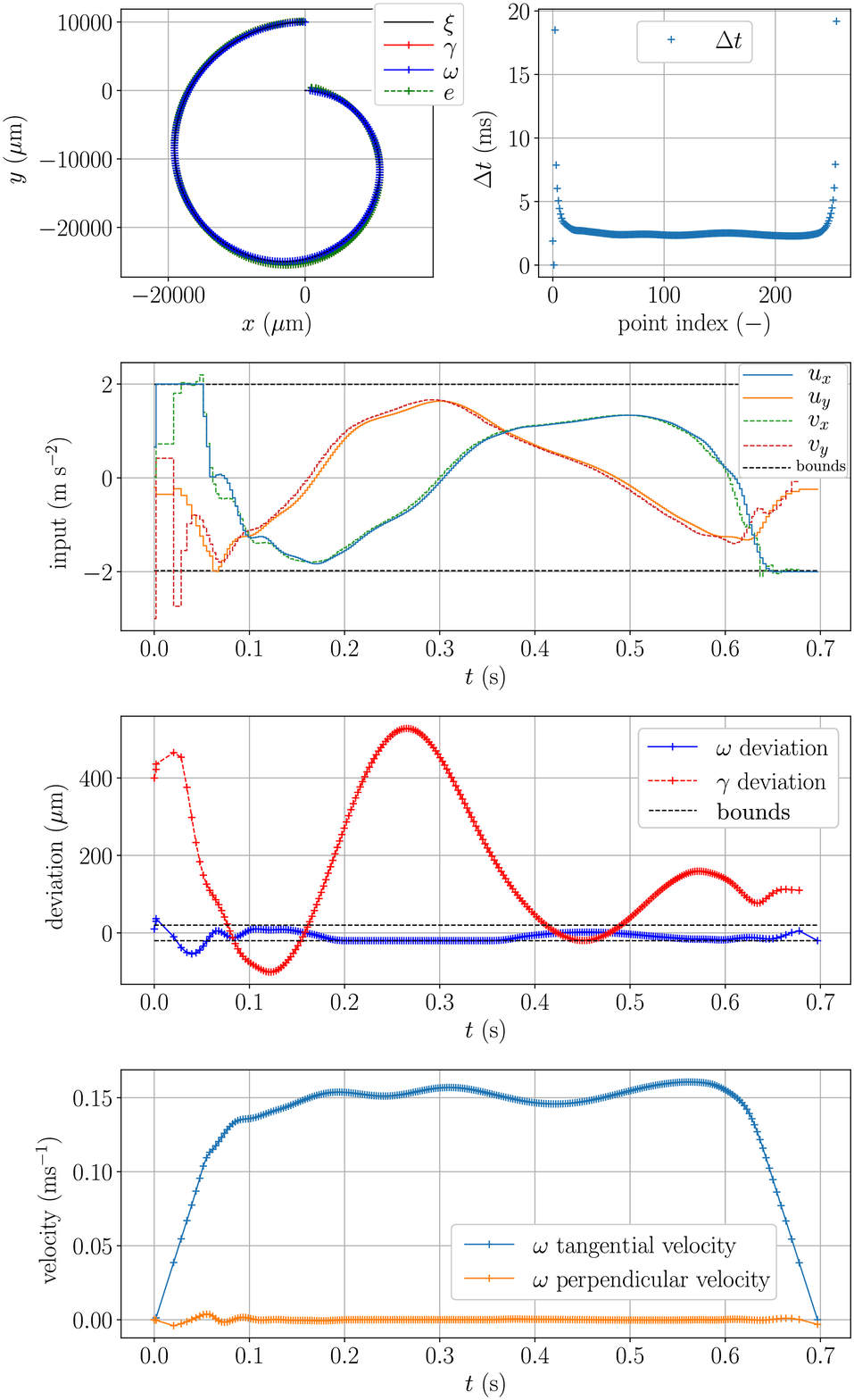} 
	  \caption{Result of the optimization for a smooth spiral, with $N=256$.
	  }
    \label{fig:rS}
  \end{center}
\vspace{-0.54cm}
\end{figure}

\begin{figure}[t]
	\vspace{3mm}
  \begin{center}
    \includegraphics[width=.92\columnwidth]{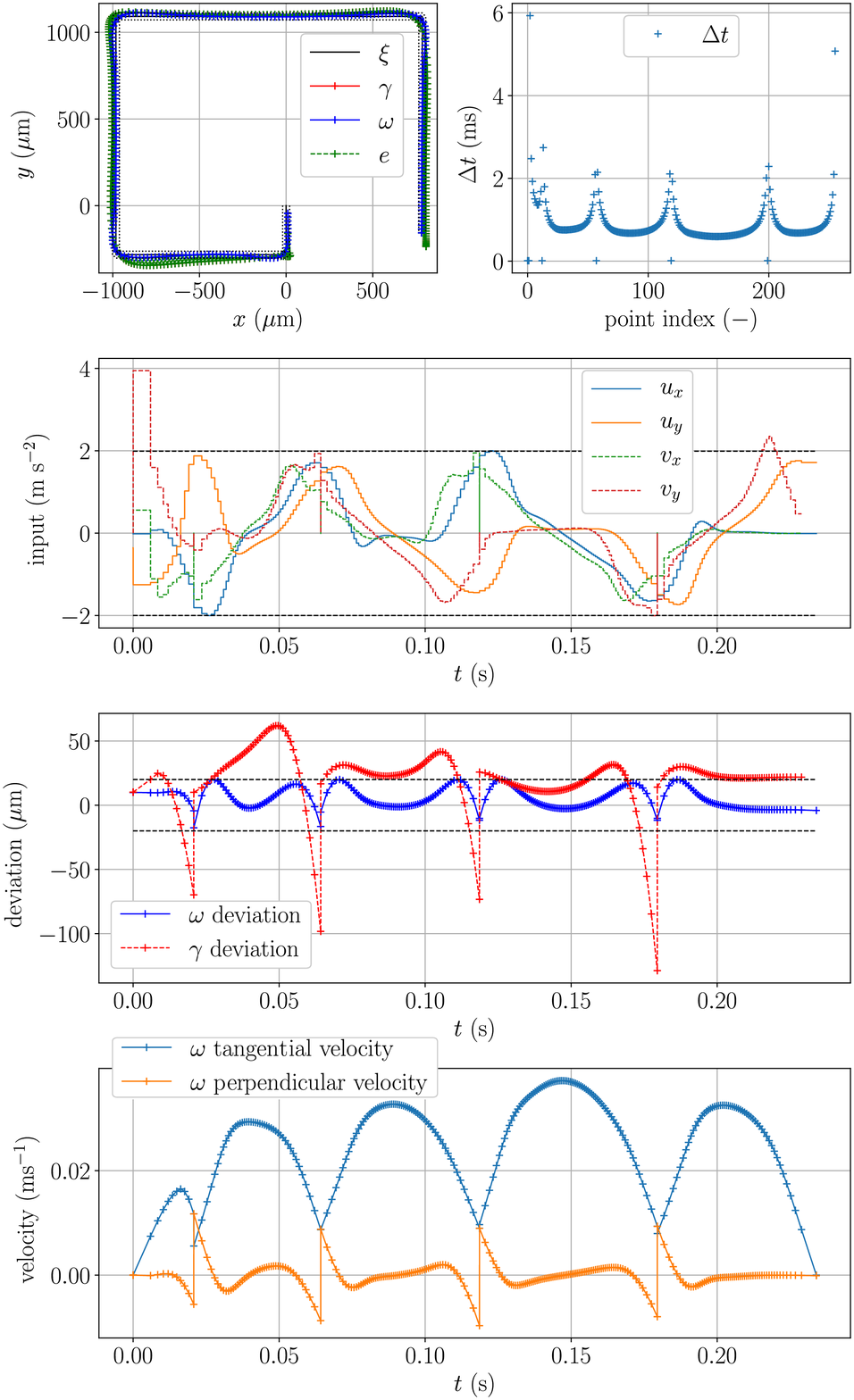} 
    \caption{Result of the optimization for a sharp spiral like shape, with $N=256$.
	  }
    \label{fig:rQ}
  \end{center}
\vspace{-0.54cm}
\end{figure}

The experimental data used to for learning the relation between inputs and tracking errors come from a different geometry than those studied in this paper, and are representative of the performance of the positioning system. We have applied the proposed method to two geometries: a Archimedean spiral with an initial radius of 10 mm and 1 turn as shown in Fig. \ref{fig:rS}, and a {\it sharp spiral} with a characteristic radius of 1 mm as shown in Fig. \ref{fig:rQ}, where the radius is a stair function with $90\deg$ steps, in order to demonstrate the performance of the approach on challenging geometries.

    \begin{figure}
		\vspace{1mm}
      \centering
    \includegraphics[width=.40\columnwidth]{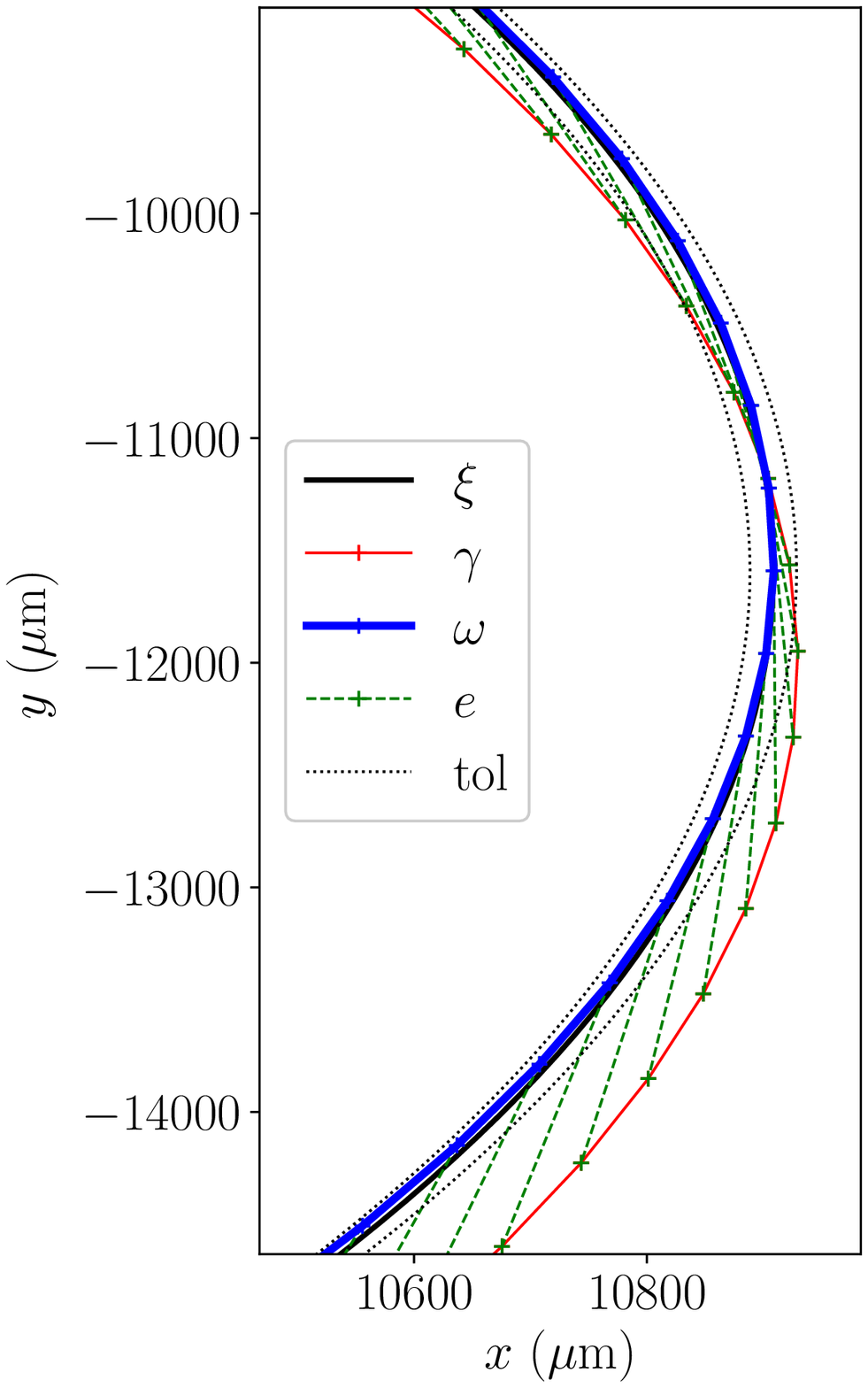}
 \label{fig:zoom_rS}
    \includegraphics[width=.40\columnwidth]{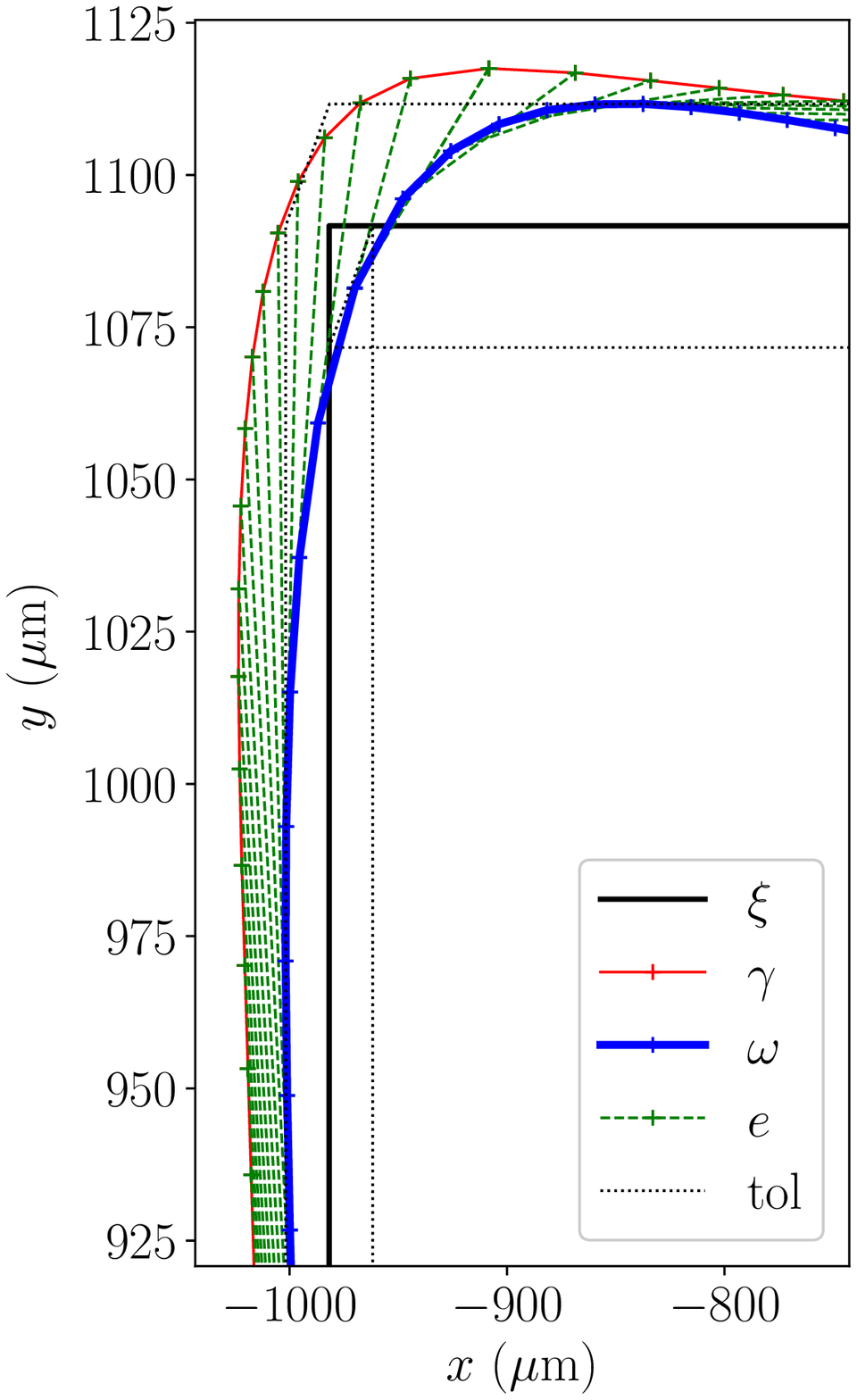}
 \label{fig:zoom_rQ}
  \caption{A close-up on the tracking performance, for the smooth (left), and sharp (right) spirals.}
  \label{fig:rsrQzoom}
    \vspace{-0.0cm}
\end{figure}
    
\begin{figure}
\centering
    \includegraphics[width=.8\columnwidth]{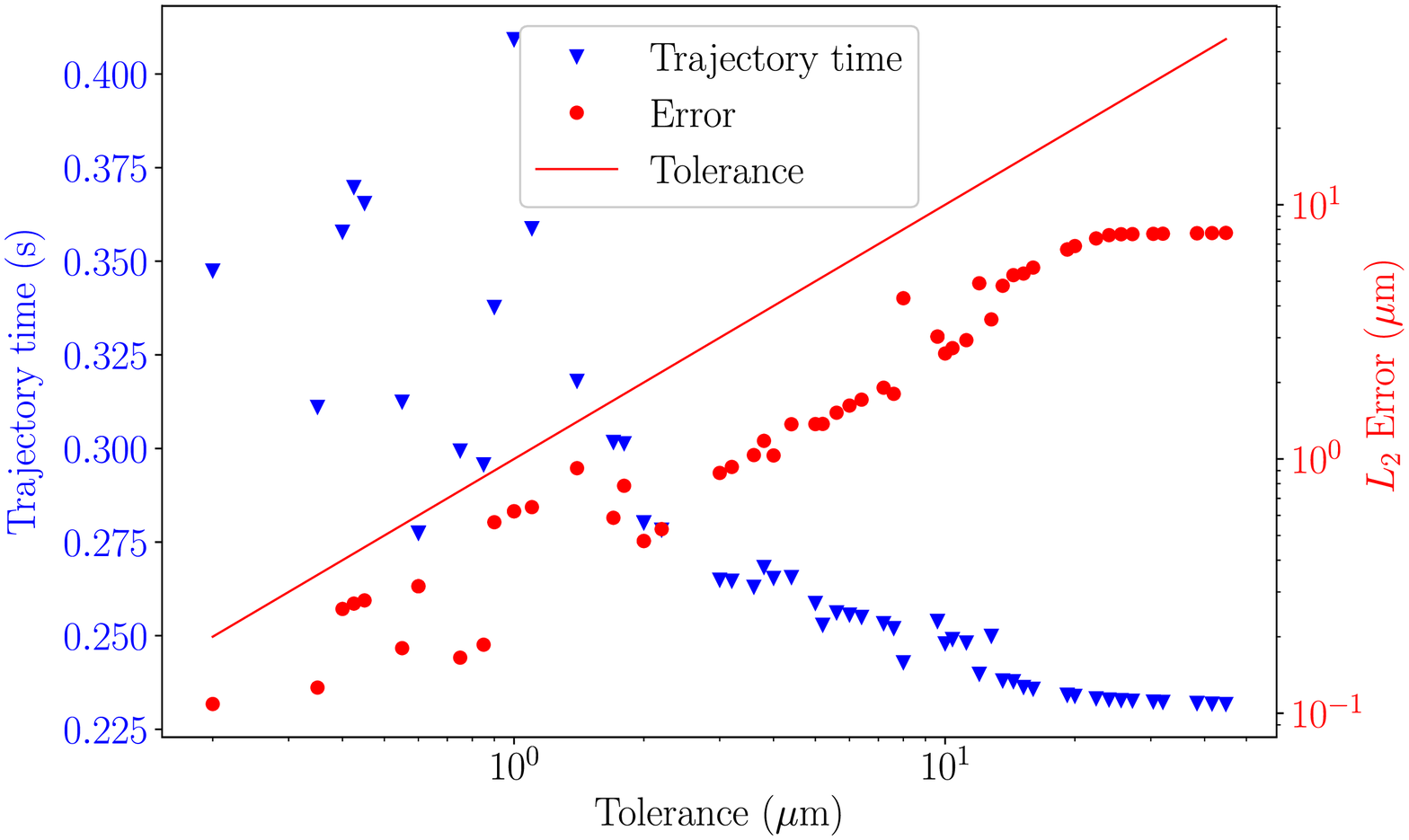}
  \caption{Trajectory time and $L_2$ error as a function of the tolerance for the sharp spiral. The red line shows the $L_2$ error if the trajectory was always at the tolerance band}
   \label{fig:ttol}
\vspace{-0.5cm}
\end{figure}

\if 01
\begin{figure}
\centering
    \includegraphics[width=.8\columnwidth]{img/xi-dev}
  \caption{Deviations from objective for the smooth spiral, where $\omega_\Gamma$ is the usual output generated from the reference $\Gamma$, and $\omega_\Xi$ is the absolute value of the output when given the objective $\Xi$ with equally spaced points in time.}
   \label{fig:xi-dev}
\vspace{-0.5cm}
\end{figure}
\fi

\if 01
\begin{table}[htbp]
\caption{Performance comparison}
\begin{center}
\begin{tabular}{|c|c|c|c|}
\hline
\textbf{Shape}&\multicolumn{3}{|c|}{\textbf{$\{t = [0, 0.01, \dots, 1], \Xi\}$}} \\
\cline{2-4} 
\textbf{}  & \textbf{\textit{time ($\mathrm{m s}$)}}& \textbf{\textit{$L_2$ error ($\mathrm{\mu m}$)}}& \textbf{\textit{$L_\infty$ error ($\mathrm{\mu m}$)}} \\
\hline
Smooth spiral   &662    &261.4  &552.1 \\ \hline
Sharp spiral    &240    &25.1   &70.0 \\
\hline
\end{tabular}
\end{center}
\vspace{.1cm}
\begin{center}
\begin{tabular}{|c|c|c|c|}
\hline
\textbf{Shape}&\multicolumn{3}{|c|}{\textbf{$\Gamma$}} \\
\cline{2-4} 
\textbf{}  & \textbf{\textit{time ($\mathrm{m s}$)}}& \textbf{\textit{$L_2$ error ($\mathrm{\mu m}$)}}& \textbf{\textit{$L_\infty$ error ($\mathrm{\mu m}$)}} \\
\hline
Smooth spiral   &662    &2.2    &9.9 \\ \hline
Sharp spiral    &240    &7.8    &20.0 \\
\hline
\end{tabular}
\label{t:xi}
\end{center}
\end{table}

\fi

The results of the optimization for a smooth spiral are shown in Fig. \ref{fig:rS}.
The objective path $\Xi$ is defined with a constant space step, with a total of $N=256$ points. Note that the definition of $\Xi$ includes only the $(x,y)$ coordinates and orientation $\alpha$, without time information. The optimal solution yields the time taken at each step. Setting the initial and final velocity to zero requires a spike in acceleration. It can be seen in the plot that the input in $x$ saturates to the bounds defined to be $u_x \in [-2, 2]$ in \eqref{bounds}. The low velocities in the beginning and end of the trajectory translate into higher time steps in these regions. The constraints in the first $16$ points are relaxed, such that even for an initial point outside the constraints, there is a feasible solution. The reference and virtual input are outside of the bounds set for the output and input since they are not required to satisfy the physical system constraints.

The output stays within the tight constraints of $20\mu m$, as defined in \eqref{bounds}. Note that the first 16 points are not constraint to lie in this zone. The tangential velocity is maximized by moving closer to the bounds, both in deviation and input. The left plot of Fig. \ref{fig:rsrQzoom} shows a close up of the achieved output.

In the presence of sharp corners as shown in Fig. \ref{fig:rQ}, the velocity has to be reduced, such that the corner can be traversed without violating the constraints.
It can be seen from the deviation subplot of Fig. \ref{fig:rQ} and from the right plot of Fig. \ref{fig:rsrQzoom} that the optimal solution takes advantage of the tolerance to cut the corners, and thus performing them at a higher speed. The velocity peaks in the mean point of the straight line segments. Note that in this case the segments are too small for the velocity to reach its maximum value. The reference and virtual input are again outside of the tolerance bounds set for the output and input since.

The effect of setting a tighter tolerance is studied in Fig. \ref{fig:ttol}. It can be seen that the time to follow the trajectory is reduced with the relaxation of the tolerance. This is to be expected, as a larger tolerance band allows to cut corners and to travel through them at higher speed. In this figure we can also see that the $L_2$ error of the trajectory increases with increased tolerance, reaching a plateau after $\approx 20\mu m$. Note that this is only the case for the particular cost function weights that have been chosen. The red line shows the $L_2$ error of the tolerance band. All feasible solutions must lie at or below this line.

\if 01
In Table \ref{t:xi} we present performance measures for the two test cases. We compare the expected output of the optimized reference $\Gamma$, and of the objective $\Xi$ assigned with constant time steps. This simulates the behavior of the machine low-level controller.
For the same trajectory time, the $L_2$ errors are reduced $100$-fold for the smooth spiral and $3$-fold for the sharp one. Note that the smooth spiral is larger, and higher velocities are achieved in this case when compared to the sharp one. The $L_\infty$ errors show that the method guarantees the satisfaction of the constraints, which are not satisfied otherwise.
\fi

\section{Conclusions}
\label{sec:conclusion}
This paper presents an approach for improving the contour control of precision machining. It is effective in designing modified reference trajectories for precision machine tools, allowing the optimization of the process without changing the internal controller, while achieving high-performance machining, with increased accuracy as compared to the nominal performance. This enables fast and low-risk retrofit, with virtually zero down time, for repeatable processes. It was tested in simulation with two types of geometries, with round or sharp corners, where the output is constrained in a $\pm 20~\mu m$ tolerance band. The proposed approach offers some advantages over the work this paper builds upon~\cite{Liniger2019}, most notably the natural way in which sharp corners are dealt with. The internal controller modelled by \eqref{eq:u-fit-lin} might need to be extended in practice, using either standard system identification techniques or more sophisticated machine learning methods. The method will be experimentally verified, and extended by automating the preprocessing of a geometry file. Another possible extension is to perform learning and correction online in a receding horizon implementation, to account for unforeseen disturbances.

\section*{Acknowledgment}
The authors would like to thank Natanael Lanz, who provided invaluable technical support on the test bench used in this work. This research has been funded by Innosuisse.

\bibliography{bib1.bib}
\bibliographystyle{IEEEtran}
\end{document}